\documentclass[referee]{raa}            

\usepackage{graphicx,times}             

\begin{document}

   \title{Analytical model of compact star in low-mass X-ray binary with de Sitter spacetime}

   \volnopage{Vol.0 (200x) No.0, 000--000}      
   \setcounter{page}{1}          

   \author{Sajahan Molla
      \inst{1}
   \and Rabiul Islam
      \inst{1}
   \and Md. Abdul Kayum Jafry
      \inst{2}
   \and  Mehedi Kalam
      \inst{1}
   }

   \institute{Department of Physics, Aliah University,
 II-A/27, New Town, Kolkata - 700156, India {\it kalam@associates.iucaa.in} \\
        \and
             Department of Physics, Shibpur Dinobundhoo Institution (College), Howrah 711102, West Bengal, India \\
   }

   \date{Received~~2009 month day; accepted~~2009~~month day}

\abstract{In this article, we introduce a de-sitter model in favor of compact stars in low-mass X-ray binary. Here, we merge the presence of the cosmological constant on a small scale to discussion the stellar structure and conclude that this doping is very well suitable with the familiar physical mode of the low-mass X-ray binary compact stars. We calculate the probable radii, compactness (u) and surface red-shift ($Z_{s}$) of six compact stars in low-mass X-ray binaries namely Cyg X-2, V395 Carinae/2S 0921-630, XTE J2123-058, X1822-371 (V691 CrA), 4U 1820-30 and GR Mus (XB 1254-690). We also offer possible equation of state (EOS) of the stellar object.
}

   \authorrunning{S. Molla, R. Islam, Md. A. K. Jafry \& M. Kalam}            
   \titlerunning{Analytical model of compact star in low-mass X-ray binary with de Sitter spacetime}  

   \maketitle

%
%
\section{Introduction}           
As compact stars (neutron stars/strange stars) plays a crucial role to relate astrophysics, nuclear physics $\&$ particle physics, it becomes a great interesting topic to study for long time. Commonly neutron stars are built almost fully of neutrons whereas strange stars are can be composed entirely of strange quark matter (SQM) or the conversion (up-down quarks to strange quarks) may be confined to the core of the neutron star (Haensel et al.~\cite{Haensel1986}; Drago et al.~\cite{Drago2014}). It is well known that neutron stars are bounded by gravitational attraction and on the other hand strange stars are bounded by strong interactions as well as gravitational attractions. Therefore, for lower mass neutron stars gravitational bound becomes much weaker than the strange stars. Hence neutron stars become larger in size in comparison to the strange stars of same mass. All the present EOS of neutron star have zero surface matter density, whereas available EOS of strange star obtained a sharp surface (Farhi \& Jaffe~\cite{Farhi1984}; Haensel et al.~\cite{Haensel1986}; Alcock et al.~\cite{Alcock1986}; Dey et al.~\cite{Dey1998}). Since within very few seconds of life of a neutron star that temperature reduced to less than Fermi energy, hence for a given equation of state the mass and radius of the star depends solely on central density and also it is very hard to find out mass and radius of a neutron star simultaneously. For a detail study we suggest a review work of Lattimer \& Prakash~(\cite{lattimer2007}). Conceptual account of mass and radii of spherically-symmetric non-rotating compact stars are results of analytical or numerical solutions of Tolman-Oppenheimer-Volkoff i.e., TOV equations. From observational point of view, some promising area for surveying of mass and radius of a compact star (neutron stars/strange stars) are thermal emission from cooling stars, pulsar timing, surface explosions and gravity wave emissions. For the experimental scientist face the recent challenges are to use giant dipole resonances, heavy-ion collisions and parity-violating electron scattering techniques to measure the density dependability pressure of nuclear matter. Actually, the most challenging task is to determine the proper EOS to describe the internal formation of a neutron star (Ozel~\cite{Ozel2006}; $\ddot{O}$zel et al.~\cite{Ozel2009a}; $\ddot{O}$zel \& Psaitis~\cite{Ozel2009b}; $\ddot{O}$zel et al.~\cite{Ozel2010}; G$\ddot{u}$ver et al.~\cite{Guver2010a}, \cite{Guver2010b}). Though a several dozen compact star masses have been determined very exactly (to some extend) in binaries (Heap \& Corcoran~\cite{Heap1992}; Van et al.~\cite{Van1995}; Stickland et al.~\cite{Stickland1997}; Orosz \& Kuulkers~\cite{Orosz1999}; Lattimer \& Prakash~\cite{lattimer2005}, \cite{lattimer2007}), no radius information is obtainable for these systems. Therefore, theoretical study of the stellar structure is essential to support the correct direction for the newly observed masses and radii. Here, some of the researcher's work on compact stars
(Lobo~\cite{Lobo2006}; Bronnikov \& Fabris~\cite{Bronnikov2006}; Hossein et al.~\cite{Hossein2012}; Rahaman et al.~\cite{Rahaman2012a}, \cite{Rahaman2012b}; Maharaj et al.~\cite{Maharaj2014}; Pant et al.~\cite{Pant2014}; Ngubelanga et al.~\cite{Ngubelanga2015}; Paul et al.~\cite{Paul2015}; Kalam et al.~\cite{Kalam2012}, \cite{Kalam2013a}, \cite{Kalam2013b}, \cite{Kalam2014a}, \cite{Kalam2014b}, \cite{Kalam2016}, \cite{Kalam2017}; Jafry et al.~\cite{Jafry2017}; Maurya et al.~\cite{Maurya2016}; Dayanandan et al.~\cite{Dayanandan2016}; Bhar et al.~\cite{Bhar2017}) are to be mentioned.

Casares et al.~(\cite{Casares2010}) surveyed the mass of the compact star in Cyg X-2 by using new high-resolution spectroscopy and it comes out as $1.71 \pm 0.21 M_{\odot}$. In another work, Steeghs \& Jonker~(\cite{Steeghs2007}) measured the mass of the compact star in V395 Carinae/2S 0921-630 with the help of MIKE echelle spectrograph on the Magellan-Clay telescope by using high-resolution optical spectroscopy and it comes out as $1.47 \pm 0.10 M_{\odot}$. On the other hand, Gelino et al.~(\cite{Gelino2003}) surveyed the mass of the compact star in XTE J2123-058 as $1.53^{+0.30}_{-0.42} M_{\odot}$.
Mu\~{n}oz-Darius et al.~(\cite{Munoz2005}) surveyed the mass of the neutron star in low-mass X-ray binary (LMXB) X1822-371  (V691 CrA)  by  perusing  the  K-correction  for the case of ejection lines formed in the X-ray illuminated atmosphere of a Roche lobe filling star and that appear as  $1.61 M_{\odot} \leq M_{NS} \leq 2.32 M_{\odot}$. In a recent work, G$\ddot{u}$ver et al.~(\cite{Guver2010b}) surveyed the mass of the compact star in 4U 1820-30 by using time resolved X-ray spectroscopy of the thermonuclear burst of 4U1820-30 and it comes out as $1.58 \pm 0.06 M_{\odot}$. Barnes et. al~(\cite{Barnes2007}) have also determined the mass of the compact object in GR Mus (XB 1254-690) as $1.20 M_{\odot} \leq M_{NS} \leq 2.64 M_{\odot}$.

Wilkinson Microwave Anisotropic Probe (WMAP) measurement indicates that in the Universe nearly 73\% of total mass-energy is Dark Energy (Perlmutter et al.~\cite{Perlmutter1998}; Riess et al.~\cite{Riess2004}) and the guidance theory of dark energy is risen on the cosmological constant, $\Lambda$ characterized by expulsive pressure which was initiated by Einstein in 1917 to achieve a static cosmological model. Later Zel'dovich (\cite{Zel'dovich1967}, \cite{Zel'dovich1968}) turned it as a vacuum energy of quantum fluctuation. However, for viability of the present-day accelerated Universe the earlier cosmological constant $\Lambda$, commonly, accepted it time-dependent in the cosmological domain (Perlmutter et al.~\cite{Perlmutter1998}; Riess et al.~\cite{Riess2004}). At the same time, space-dependent $\Lambda$ has an desired outcome in the astrophysical point of view as argued by other researchers (Chen \& Wu~\cite{Chen1990}; Narlikar et al.~\cite{Narlikar1991}; Ray \& Ray~\cite{Ray1993}) in respect to the behaviour of local massive objects kind of galaxies and elsewhere. In the present motto of compact stars, however, we take cosmological constant, $\Lambda$ as a absolutely constant quantity. This constancy of $\Lambda$ unable to ruled out for the scheme of very small dimension like as compact star systems or elsewhere with various physical needs (MaK~\cite{MaK2000}; Dymnikova~\cite{Dymnikova2002}; Dymnikova~\cite{Dymnikova2003}; B$\ddot{O}$hmer \& Harko~\cite{Bohmer2005}). To estimate mass and radii regarding neutron star Egeland (\cite{Egeland2007}) incorporated the presence of cosmological constant proportionality trust on the density of vacuum. Egeland have done it by application the Fermi equation of state along with the Tolman-Oppenheimer-Volkoff (TOV) equation.

Voluntary by the above knowledge, we organize the presence of cosmological constant in a small scale to exercise the construction of compact stars in low-mass X-ray binaries namely Cyg X-2, V395 Carinae/2S 0921-630, XTE J2123-058, X1822-371 (V691 CrA), 4U 1820-30 and GR Mus (XB 1254-690) and attained to a finality that incorporation of $\Lambda$ tells the compact stars in good manners.

\section{Interior Spacetime}
We consider stars as static and spherically symmetric body whose interior spacetime as
\begin{equation}
ds^2=-e^{\nu(r)}dt^2 + e^{\lambda(r)}dr^2 +r^2(d\theta^2 +sin^2\theta d\phi^2) \label{eq1}
\end{equation}
According to Heintzmann~ (\cite{Heintzmann1969}),
\begin{eqnarray}
e^{\nu} &=&
A^2\left(1+ar^2\right)^{3} \nonumber
\end{eqnarray}
and
\begin{eqnarray}
e^{-\lambda} &=&
\left[1-\frac{3ar^2}{2} \left(\frac{1+C\left(1+4ar^2\right)^{-\frac{1}{2}}}{1+ar^2}\right) \right] \nonumber
\end{eqnarray}
where A, C and $a$ are constants. \\
We presume that the energy-momentum tensor for the interior of the compact object has the standard form as
\[T_{ij}=diag(-\rho,p,p,p)\]
where $\rho$ and $p$ are energy density and isotropic pressure respectively.\\
Einstein's field equations for the metric equation (1) in presence of $\Lambda$ are then obtained as (taking $G =1$ and $ c = 1$)
\begin{eqnarray}
\label{eq2}
 8\pi\rho+ \Lambda &=&
e^{-\lambda}\left(\frac{\lambda^\prime}{r}-\frac{1}{r^2}\right) +
\frac{1}{r^2}\\
\label{eq3}
8\pi p - \Lambda &=&
e^{-\lambda}\left(\frac{\nu^\prime}{r}+\frac{1}{r^2}\right) -
\frac{1}{r^2}
\label{eq4}
\end{eqnarray}

Now, from the metric equation (1) and the Einstein's field equations (2) \& (3), we obtain the energy density ($\rho$) and the pressure ($p$) as

\begin{eqnarray}
\rho &=& \frac{3a\left(\sqrt{1+4ar^2}\left(3+13ar^2+4a^2r^4\right)+C\left(3+9ar^2\right)\right)}{16\pi\left(1+ar^2\right)^2\left(1+4ar^2\right)^{\frac{3}{2}}}
-\frac{\Lambda}{8\pi}
\end{eqnarray}

\begin{eqnarray}
p &=& \frac{-3a\left(3\sqrt{1+4ar^2}\left(-1+ar^2\right)+C+7aCr^2\right)}{16\pi\left(1+ar^2\right)^2\left(1+4ar^2\right)^{\frac{1}{2}}}
 +\frac{\Lambda}{8\pi}
\end{eqnarray}

From the equation (4) and equation (5) we get the central density ($ \rho_0$) and central pressure ($ p_0$) of the star gradually:
\begin{eqnarray}
\rho_0 = \rho(r=0) = \frac{3a\left(3+3C\right)}{16\pi}-\frac{\Lambda}{8\pi}  \nonumber\\
p_0 = p(r=0) = \frac{3a\left(3-C\right)}{16\pi}+\frac{\Lambda}{8\pi}  \nonumber
\end{eqnarray}

\begin{figure}[htbp]
\centering
\includegraphics[scale=.25]{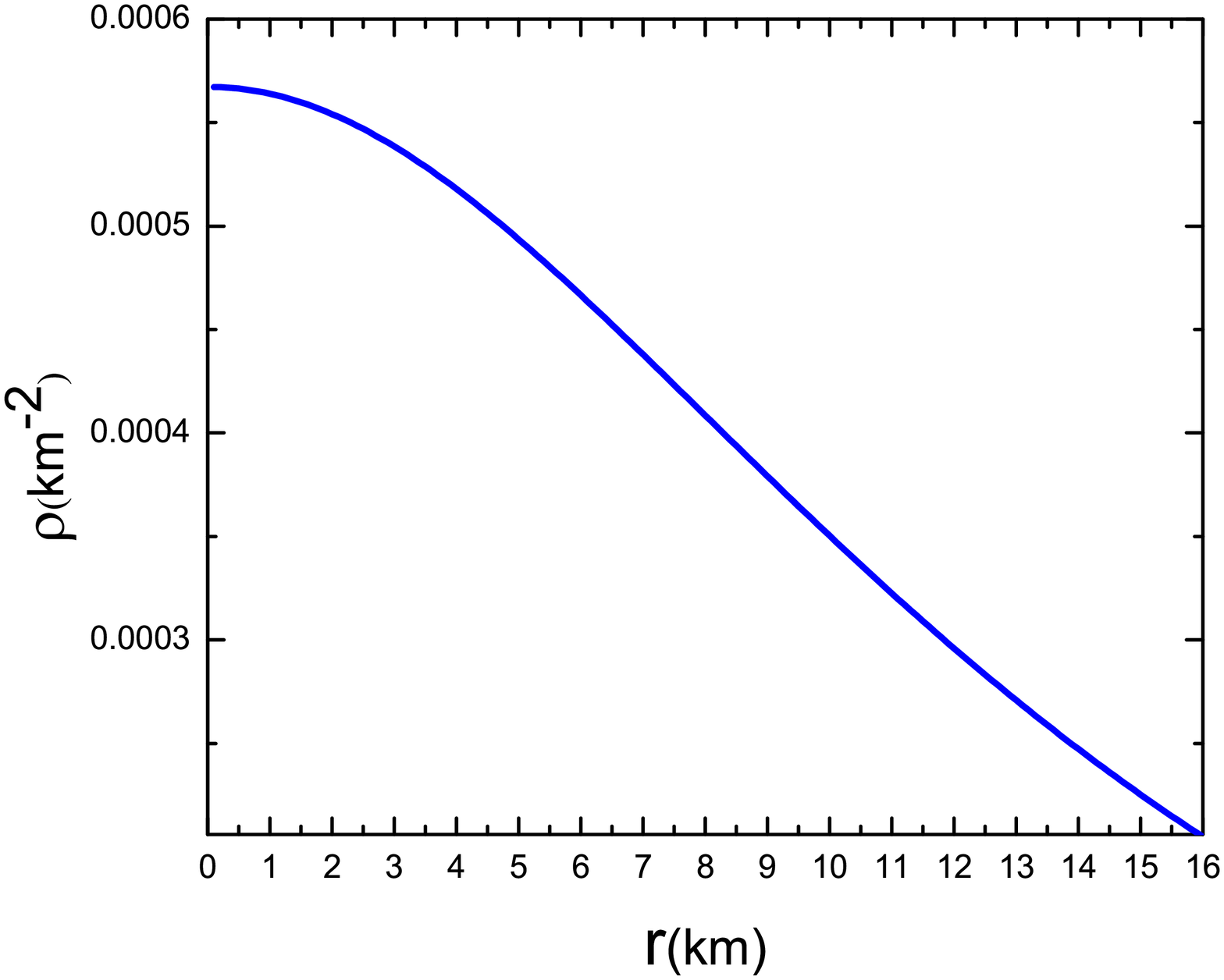}
\includegraphics[scale=.25]{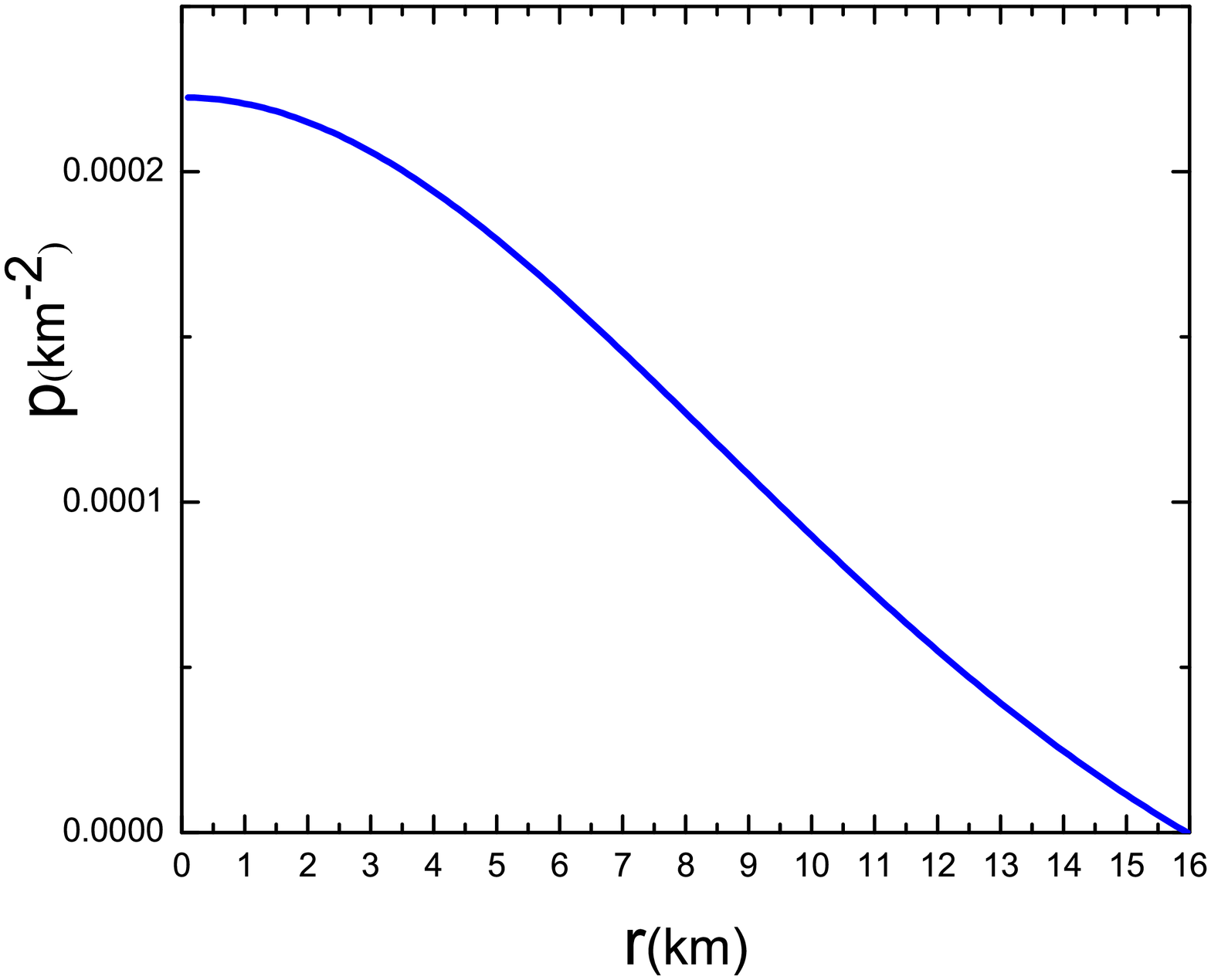}
\caption{Variation of Matter density ($\rho$) - radius (r) and pressure ($p$) - radius (r) at the stellar interior (taking a=0.0016 $km^{-2}$, C=1.133).}
\label{fig:1}
\end{figure}

It is known that, $\Lambda>0$ suggested the space is open. In order to interpret the nowadays acceleration state of the universe, it has been established that energy in the vacuum is liable for this expansion. As a consequence, vacuum energy give some gravitational influence on the stellar structures. It is recommended that cosmological constant plays the preface of energy of the vacuum. The value of cosmological constant, $\Lambda$ has not been consistent with various scenarios. Though in the cosmological point of view, its order of magnitude possibly roughly $10^{-52}m^{-2}$, in the local scale (for example near the black holes, neutron stars and various massive objects) it is not essential to follow the large scale fine tuning values of $\Lambda$ (Bordbar et al.~\cite{Bordbar2016}).

In this part we will perusal the following features of our model presuming the value of $\Lambda =0.00111 km^{-2}$ (nearer to the value of B$\ddot{O}$hmer \& Harko~\cite{Bohmer2005a}; Bordbar et al.~\cite{Bordbar2016}). We have considered this value for the mathematical consistency and stability of the compact star. As “$a$” and “$C$” specify the central density of the configurations, we calculate it and use it in our model as we know that inward properties of the compact star depends on the central density.

Also, we observe (Fig.~1) that, matter density and pressure both are maximum at the centre and decreases monotonically unto the boundary. Interestingly, pressure fall to zero at the boundary, though density does not. Therefore, it may be justified to take these compact stars as a strange stars where the surface density remains finite rather than the neutron stars for which the surface density vanishes at the boundary (Farhi \& Jaffe~\cite{Farhi1984}; Haensel et al.~\cite{Haensel1986}; Alcock et al.~\cite{Alcock1986}; Dey et al.~\cite{Dey1998}). It is to be mentioned here that, we fittings the values of the constants  $a=0.0016 km^{-2}$and $ C =1.133$, like that the pressure fall from its maximum value (at centre) towards zero at the boundary.

\section{Exploration of Physical properties}
In this section we examine the following property of the compact star in low-mass X-ray binary:

\subsection{Energy conditions}
In our model we observed that all the energy conditions, namely null energy condition(NEC), weak energy condition(WEC), strong energy condition(SEC)
and dominant energy condition(DEC) are satisfied at the centre ($r=0$) of the star. From Fig.~1, we observe that all the energy conditions maintain well:\\
(i) NEC: $p_{0}+\rho_{0}\geq0$ ,\\
(ii) WEC: $p_{0}+\rho_{0}\geq0$  , $~~\rho_{0}\geq0$  ,\\
(iii) SEC: $p_{0}+\rho_{0}\geq0$  ,$~~~~3p_{0}+\rho_{0}\geq0$ ,\\
(iv) DEC: $\rho_{0} > |p_{0}| $.

See Table~1 for numerical justification of energy conditions satisfied in our model.
\begin{table*}[h]
\centering
\caption{Evaluated parameters for energy conditions in our model receiving $a$=0.0016 $km^{-2}$, C=1.133.}
\begin{tabular}{@{}cccc@{}} 
\hline
 $\rho_{0}$ (km$^{-2}$) & $p_{0}$ (km$^{-2}$) & $\rho_{0}$+$p_{0}$ (km$^{-2}$)  & 3$p_{0}$+$\rho_{0}$ (km$^{-2}$)  \\
\hline
 0.000566894 & 0.000222451 & 0.000789345 & 0.00123425 \\ 
\hline
\end{tabular}
\end{table*}

\subsection{TOV equation}
In our stellar model we observe that static equilibrium configurations attend due to gravitational ($F_g$) and hydrostatic ($F_h$) forces presence.
\begin{eqnarray}
F_h+ F_g  = 0 \nonumber
\end{eqnarray}
where,
\begin{eqnarray}
F_g &=& \frac{1}{2} \nu^\prime\left(\rho+p\right) \nonumber \\
F_h &=& \frac{d}{dr}(p-\frac{\Lambda}{8\pi}) \nonumber
\end{eqnarray}
Fig.~2 shows that equilibrium state of the compact object under gravitational and hydrostatic forces in our stellar model.

\begin{figure}[htbp]
\centering
\includegraphics[scale=.3]{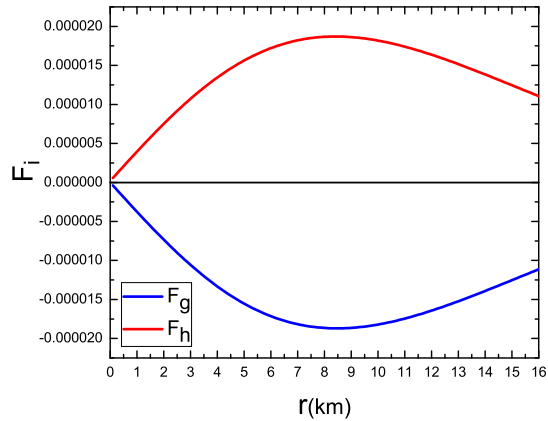}
\caption{Behaviors of the gravitational ($F_g$) and hydrostatic ($F_h$) forces at the stellar inside.}
\label{fig:2}
\end{figure}

\subsection{Stability}
For a stable stellar model it is always required that the speed of sound should be less than speed of light $(c=1)$ everywhere within the stellar object i.e. $0 \leq  v^2=(\frac{dp}{d\rho})\leq 1$ (Herrera~\cite{Herrera1992}; Abreu et al.~\cite{Abreu2007}). For these purpose we plot the sound speed in Fig.~3(left) and
observed that it satisfies well the inequalities $0\leq v^2 \leq 1$.
Therefore our stellar model is well stabled.

Our stellar model is also dynamical stable in present of thermal radiation. The dynamical stability examined by adiabatic index ($\gamma$). The adiabatic index ($\gamma$) is identify as
\begin{eqnarray}
 \gamma = \frac{\rho+p}{p} \frac{dp}{d\rho} \nonumber
\end{eqnarray}
If the value of adiabatic index $\gamma > \frac{4}{3}$ through out the interior of the stellar body then the stellar model will be stable.
From Fig.~3(right) we observe that our stellar model is stable in present of thermal radiation. This type of stability executed by several author namely Chandrasekhar~(\cite{Chandrasekhar1964}), Bardeen et al.~(\cite{Bardeen1966}), Knutsen~(\cite{Knutsen1988}), Mak \& Harko~(\cite{Mak2013}) gradually
in their work.

\begin{figure}[htbp]
\centering
\includegraphics[scale=.25]{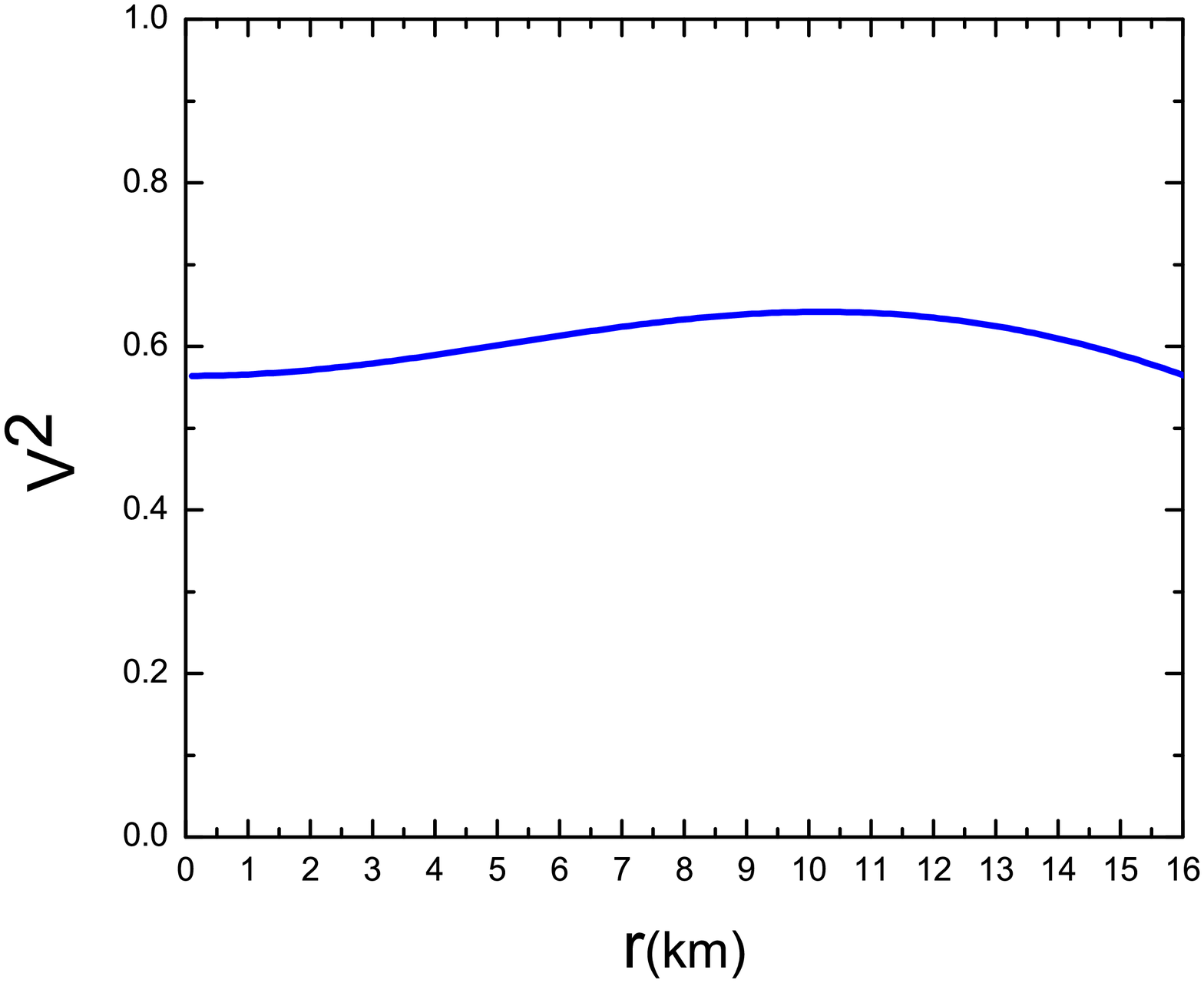}
\includegraphics[scale=.25]{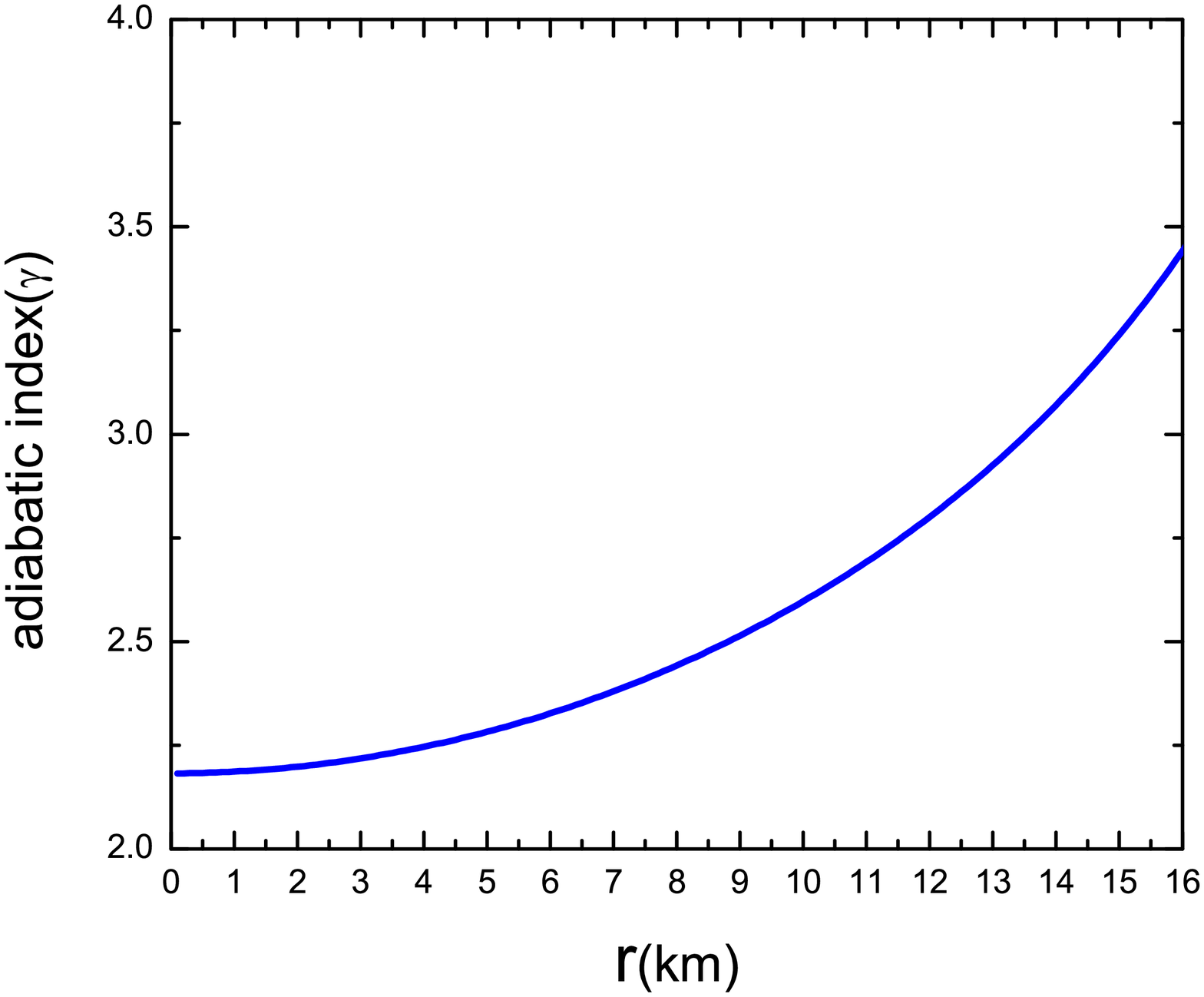}
\caption{Sound speed ($v^2$) - radius ($r$) and adiabatic index ($\gamma$) - radius ($r$) relation at the stellar inside (taking a=0.0016 $km^{-2}$, C=1.133).}
\label{fig:4}
\end{figure}

\subsection{Matching Conditions}
Here, we match the interior metric of the star with the exterior Schwarzschild de Sitter metric at the boundary (r = b)
\begin{eqnarray}
ds^{2}=-\left(1 - \frac{2M}{r}-\frac{\Lambda r^2}{3} \right)dt^2 + \left(1 - \frac{2M}{r}-\frac{\Lambda r^2}{3} \right)^{-1}dr^2 + r^2(d\theta^2+\sin^2\theta d\phi^2)
\end{eqnarray}
From the continuity of the metric function across the boundary, we get the compactification factor as
\begin{equation}
\frac{M}{b} =  \frac{1}{2}[\frac{3ab^2\left(1+C(1+4ab^2)^{-\frac{1}{2}}\right)}{2\left(1+ab^2\right)}-\frac{\Lambda b^2}{3}]
\end{equation}

\subsection{Mass-Radius relation and Surface redshift}
For a static spherically symmetric perfect fluid sphere maximum allowable mass-radius ratio should be $\frac{ Mass}{Radius} < \frac{4}{9}$ (Buchdahl~\cite{Buchdahl1959}). In our stellar model we have calculated the gravitational mass (M) in presence of cosmological constant as
\begin{equation}
\label{eq34}
 M=4\pi\int^{b}_{0} \rho~~ r^2 dr = \frac{3ab^3\left(1+C(1+4ab^2)^{-\frac{1}{2}}\right)}{4\left(1+ab^2\right)}-\frac{\Lambda b^3}{6}
\end{equation}
where the radius of the star is taken $b$.\\
Hence, the compactness (u) of the star be able to written as
\begin{equation}
u = \frac{M}{b} =  \frac{1}{2}[\frac{3ab^2\left(1+C(1+4ab^2)^{-\frac{1}{2}}\right)}{2\left(1+ab^2\right)}-\frac{\Lambda b^2}{3}]
\end{equation}
The behaviour of Mass function and Compactness of the star in our model are shown in Fig.~4 and Fig.~5(left).\\
The surface redshift ($Z_s$) analogous to the above compactness ($u$) is obtained as
\begin{equation}
 Z_s= \left[ 1-2u \right]^{-\frac{1}{2}}-1
\end{equation}
Therefore the maximum surface redshift, from Fig.~5(right) for the different compact stars can be easily find out. The radii, compactness and surface redshift of the different compact stars are obtained from Fig.~6, equation (9) and equation (10) and a comparative analysis has been done in Table~2.

\begin{figure}[htbp]
\centering
\includegraphics[scale=.3]{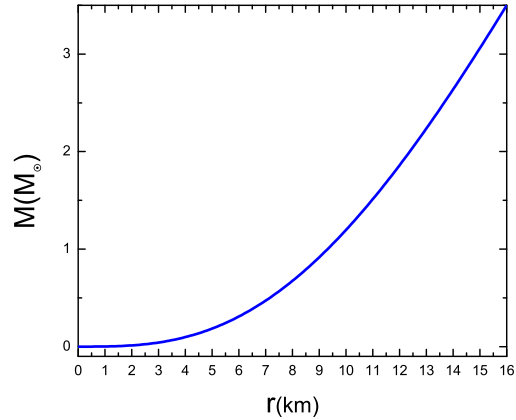}
\caption{Variation of the mass function M(r) of our star model (taking a=0.0016 $km^{-2}$, C=1.133).} \label{fig:4}
\end{figure}
\begin{figure}[htbp]
\centering
\includegraphics[scale=.25]{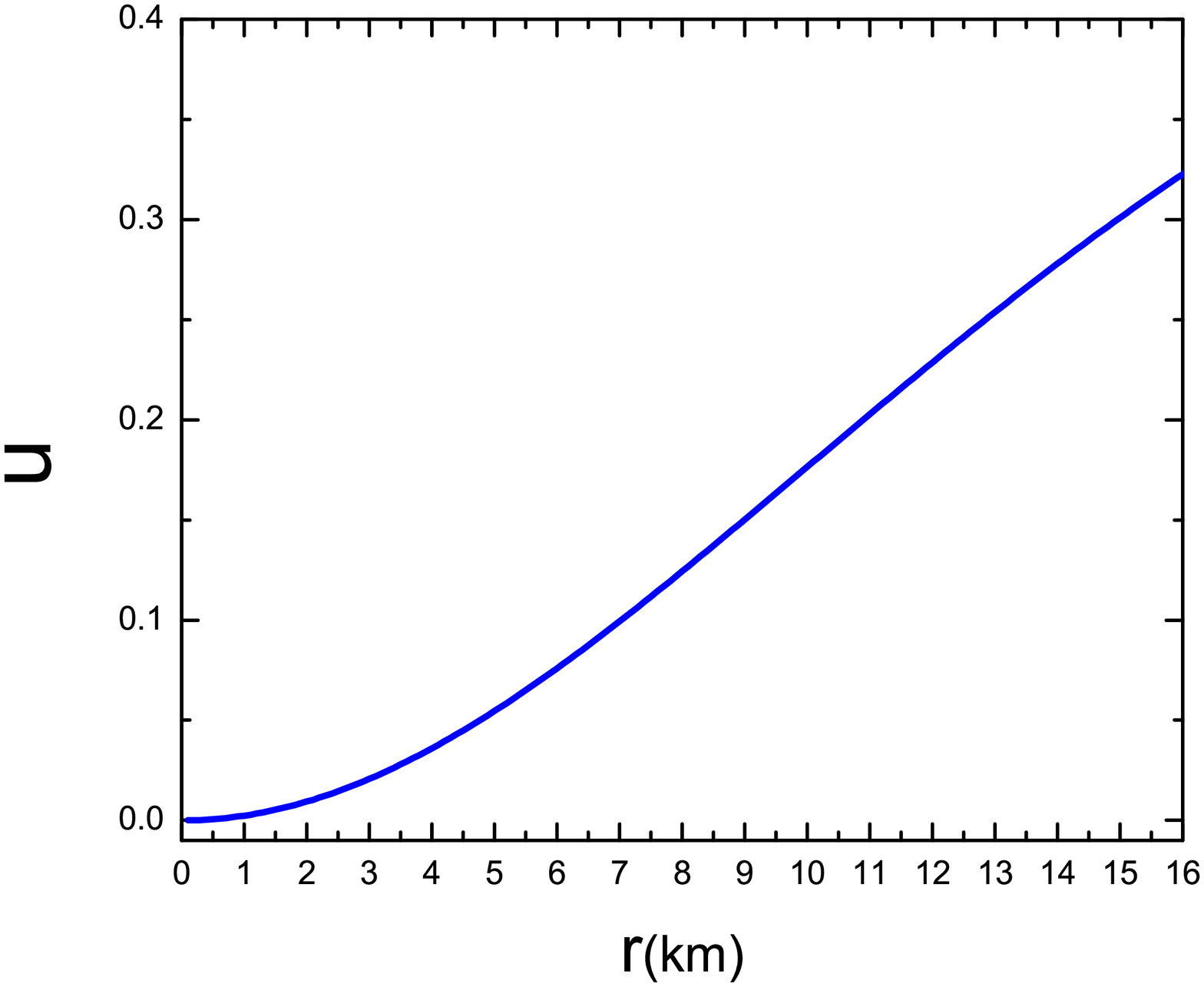}
\includegraphics[scale=.25]{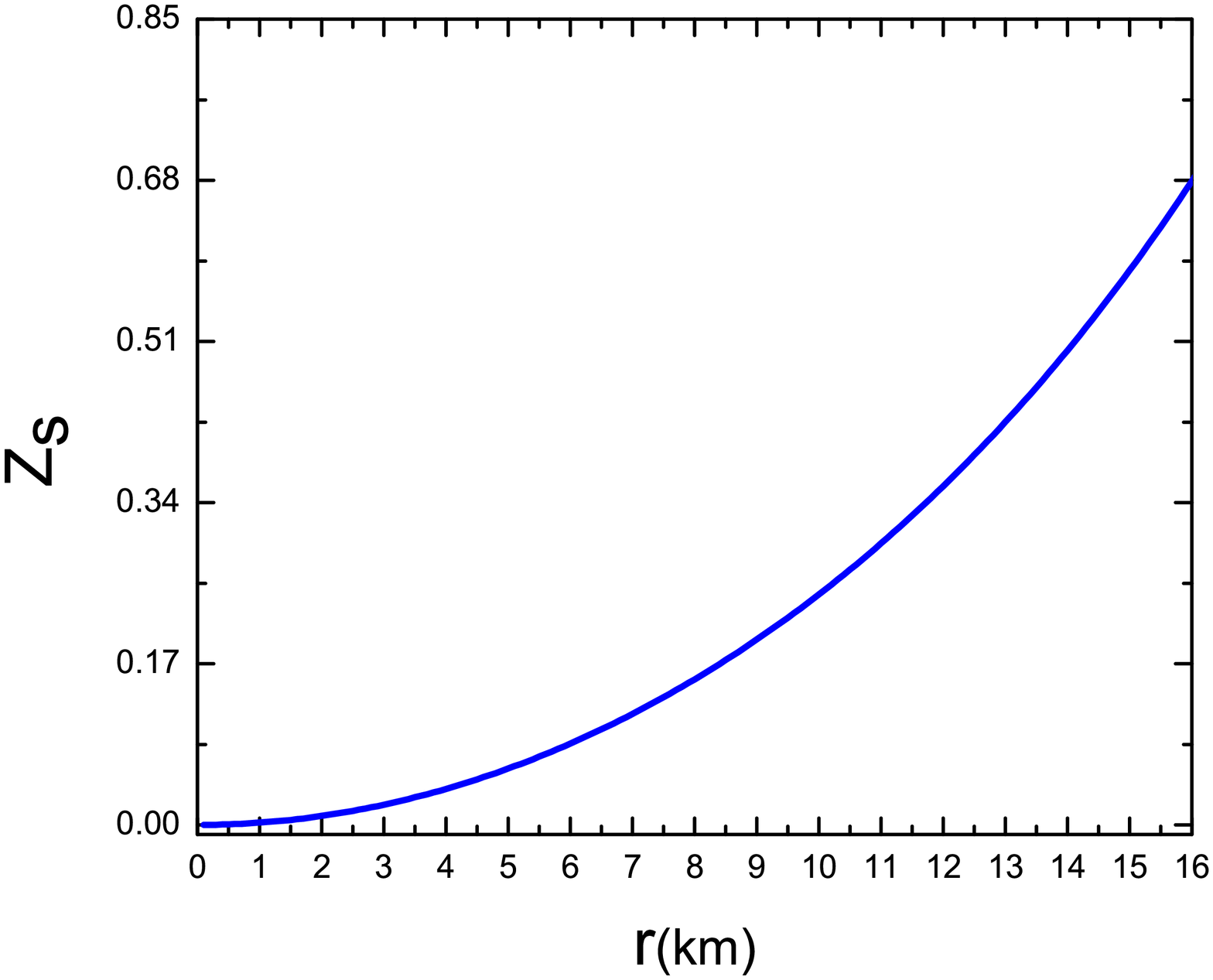}
\caption{Variation of the compactness (u) and surface red-shift ($Z_s$) of our star model (taking a=0.0016 $km^{-2}$, C=1.133).} \label{fig:5}
\end{figure}

\begin{figure}[htbp]
\centering
\includegraphics[scale=.25]{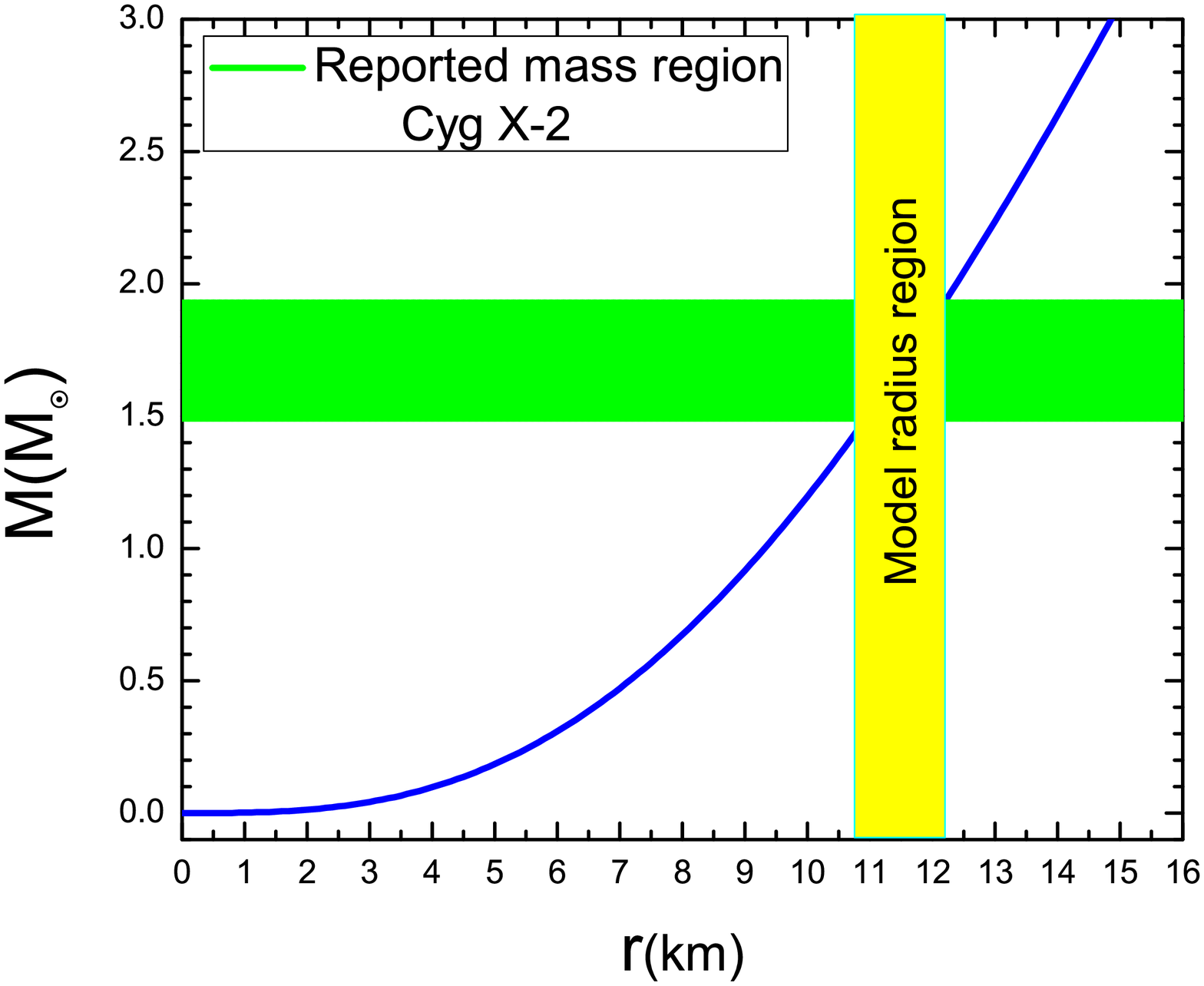}
\includegraphics[scale=.25]{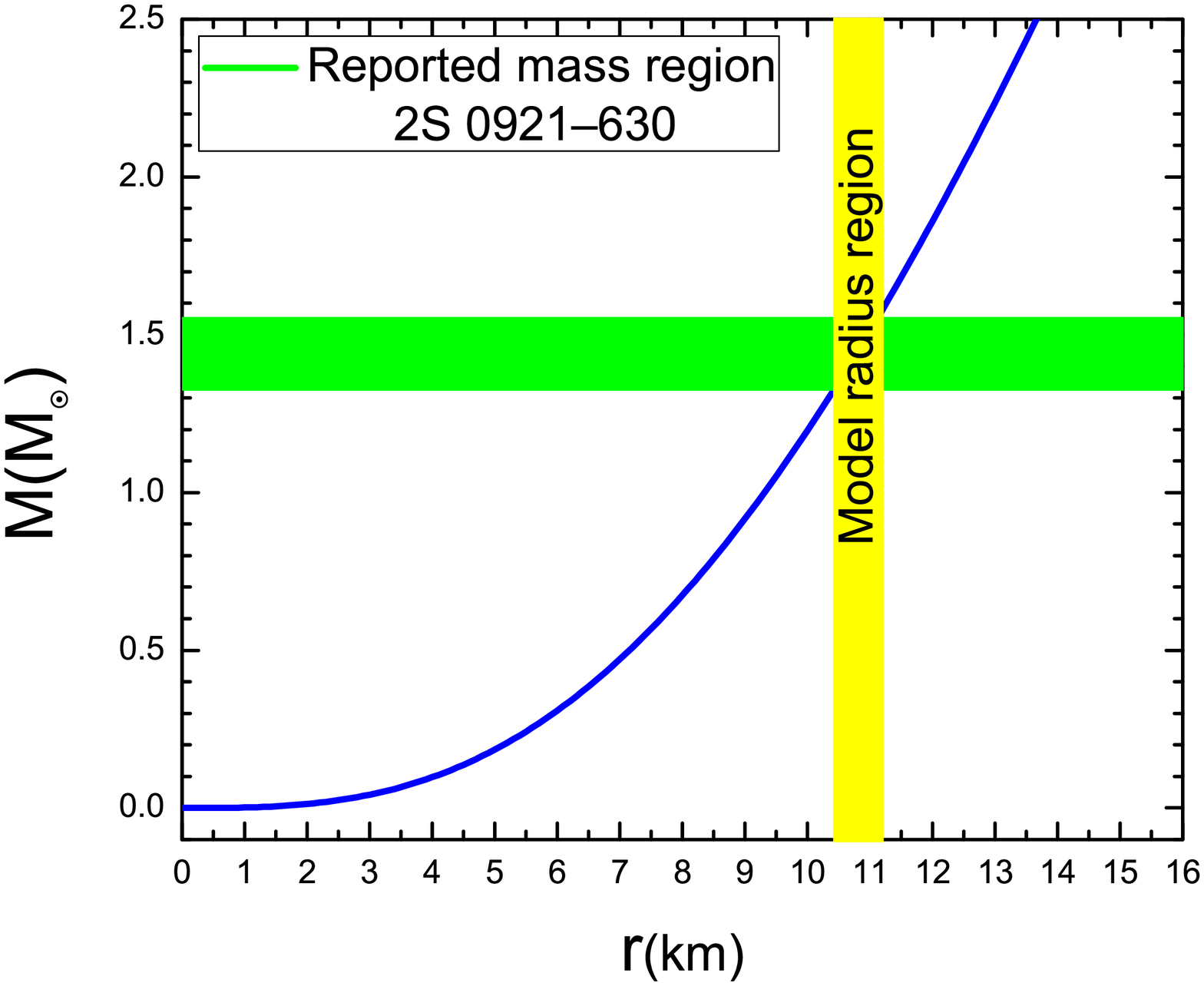}
\includegraphics[scale=.25]{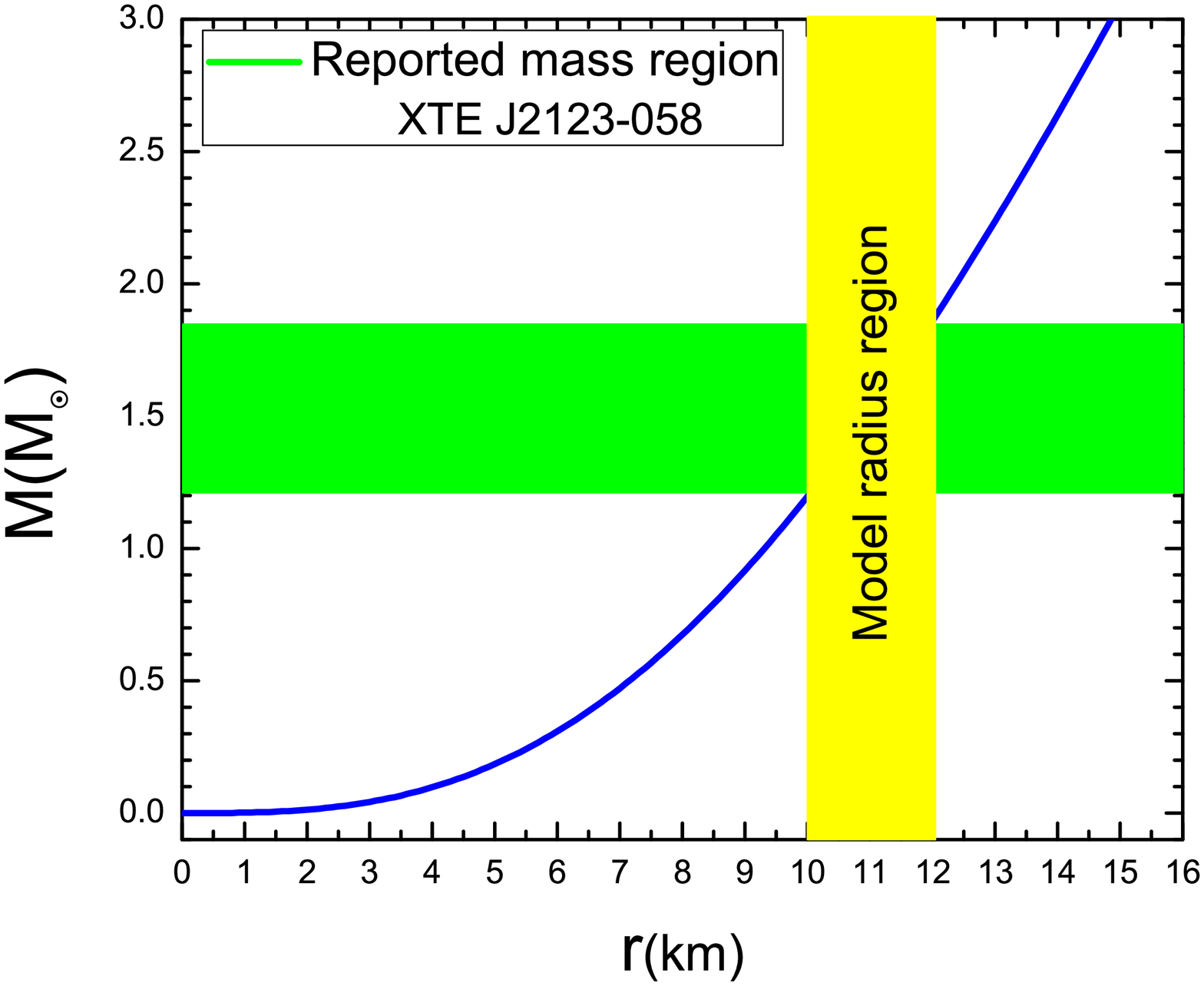}
\includegraphics[scale=.25]{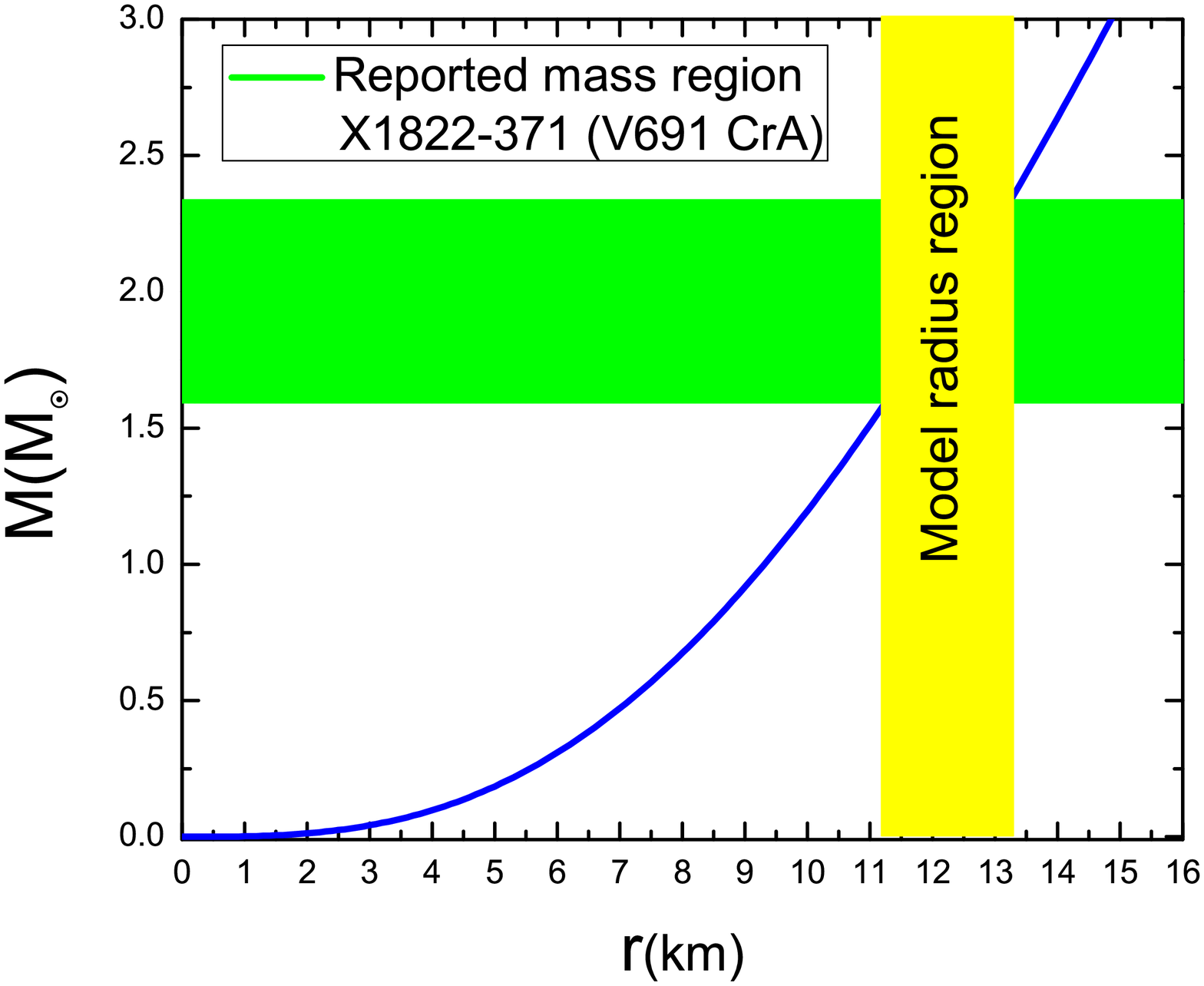}
\includegraphics[scale=.25]{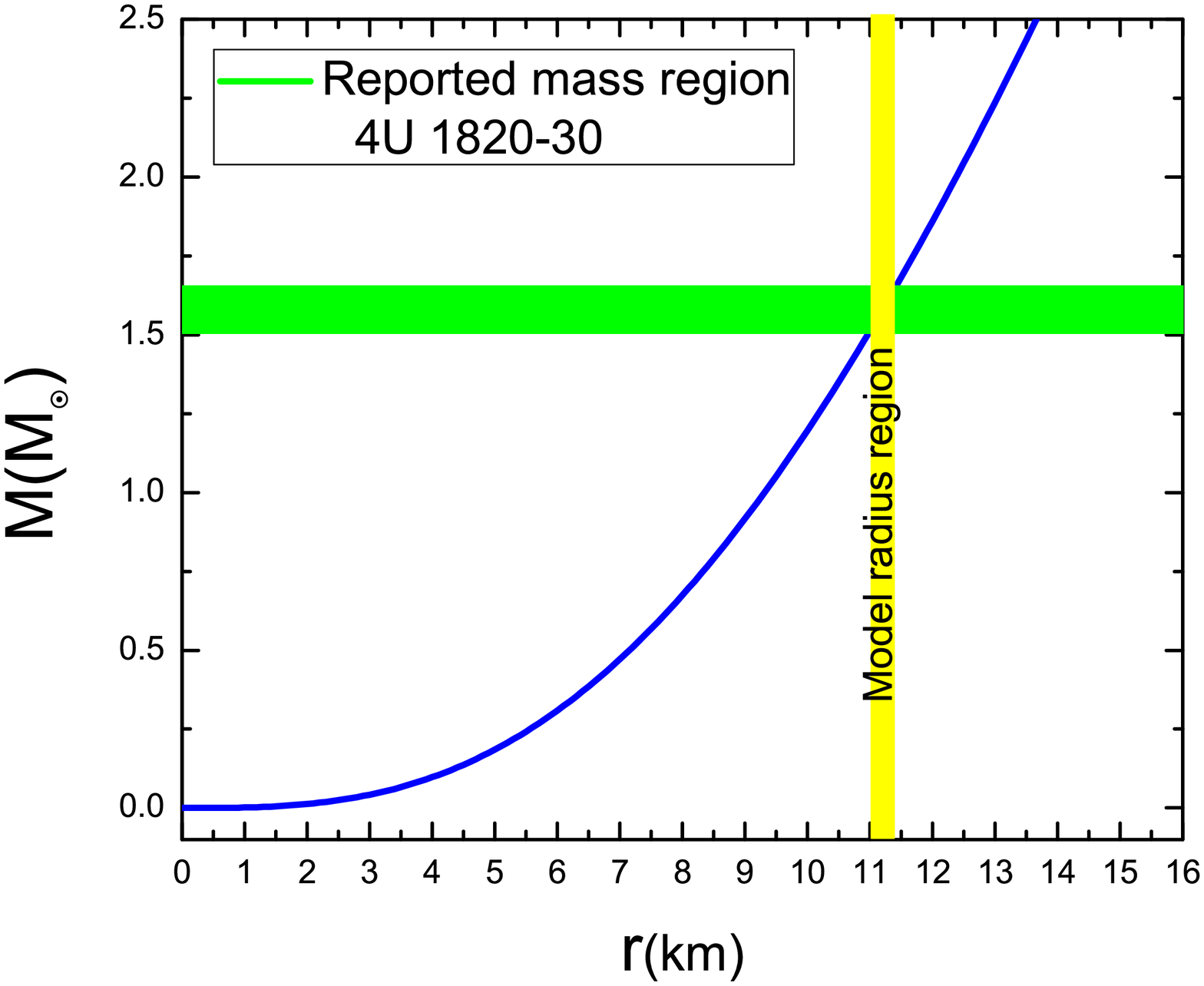}
\includegraphics[scale=.25]{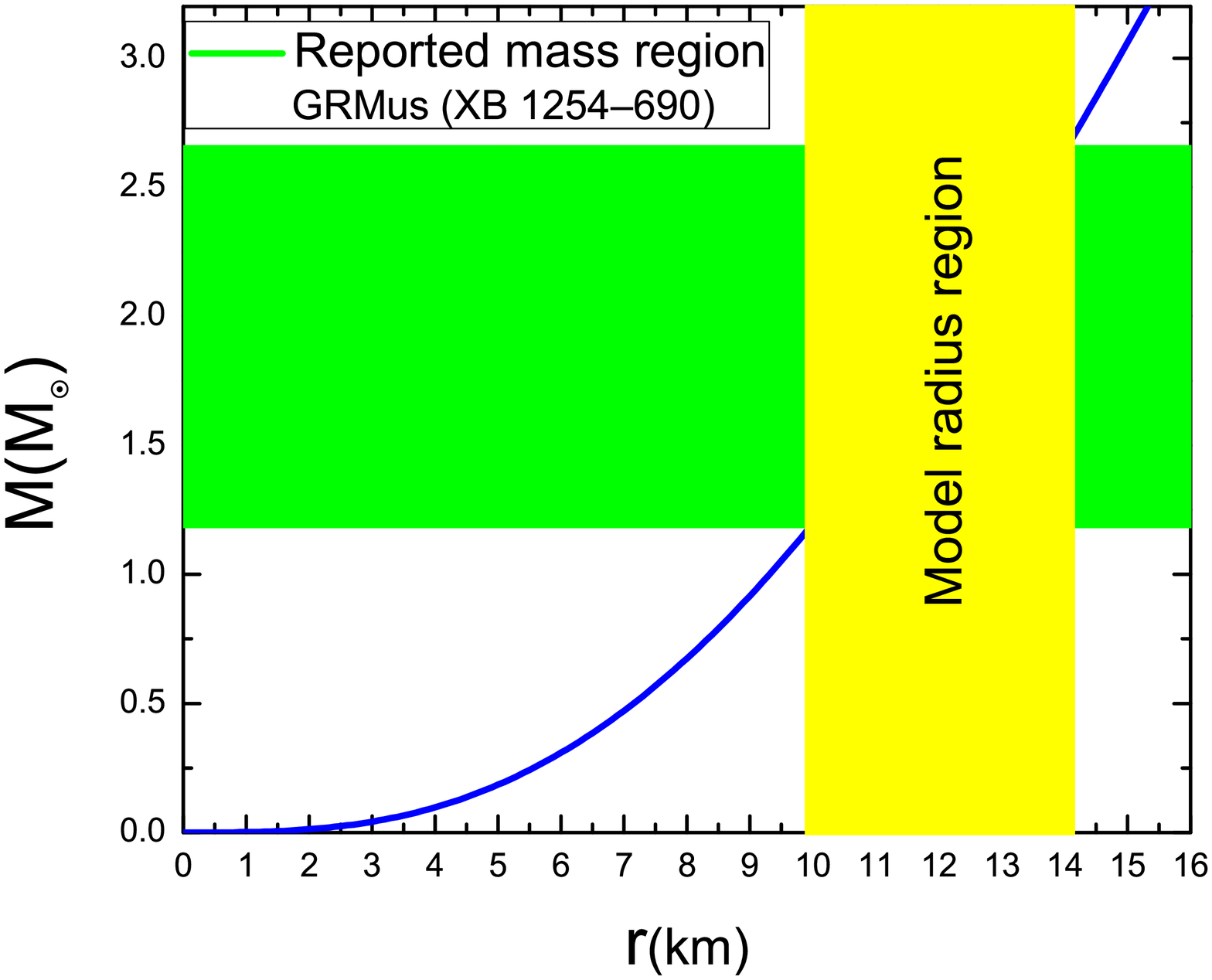}
\caption{Probable radii of Cyg X-2, 2S 0921-630, XTE J2123-058, X1822-371 (V691 CrA), 4U 1820-30 and GR Mus (XB 1254-690)  .}
\label{fig:4}
\end{figure}

\begin{table*}[h]
\centering
\caption{Evaluated parameters for compact Stars.}
\scalebox{0.90}{
\begin{tabular}{@{}ccccc@{}} 
\hline
 Star & Observed Mass($M_{\odot}$) & Radius from Model(in km) & Redshift from Model & Compactness from Model \\
\hline
 Cyg X-2 & 1.71 $\pm$ 0.21 & 11.55 $\pm$ 0.65      & 0.331 $\pm$ 0.0393 & 0.2169 $\pm$ 0.017\\
  2S 0921-630 & 1.44 $\pm$ 0.10 & 10.75 $\pm$ 0.35 & 0.2834 $\pm$ 0.0194 & 0.1962 $\pm$ 0.0092\\
 XTE J2123-058 & $1.53^{+0.30}_{-0.42}$ & 10.8 $\pm$ 1.1  & 0.2897 $\pm$ 0.0614 & 0.1973 $\pm$ 0.0287\\
 X1822-371 (V691 CrA) & 1.61$\leq$ M $\leq$ 2.32 & 11.2$\leq$ R $\leq$ 13.2 & 0.3740 $\pm$ 0.0655 & 0.2334 $\pm$ 0.0254 \\
  4U 1820-30 & 1.58 $\pm$ 0.06 & 11.2 $\pm$ 0.2 & 0.3087 $\pm$ 0.0117 & 0.208 $\pm$ 0.0052\\
  GR Mus (XB 1254-690) & 1.92 $\pm$ 0.72 & 12 $\pm$ 2.1 & 0.3734 $\pm$ 0.1352 & 0.2285 $\pm$ 0.0547\\ 
 \hline
 \end{tabular}}
\end{table*}

\section{Discussion and Concluding Remarks}
\label{sec:3}
\begin{figure}[htbp]
    \centering
        \includegraphics[scale=.3]{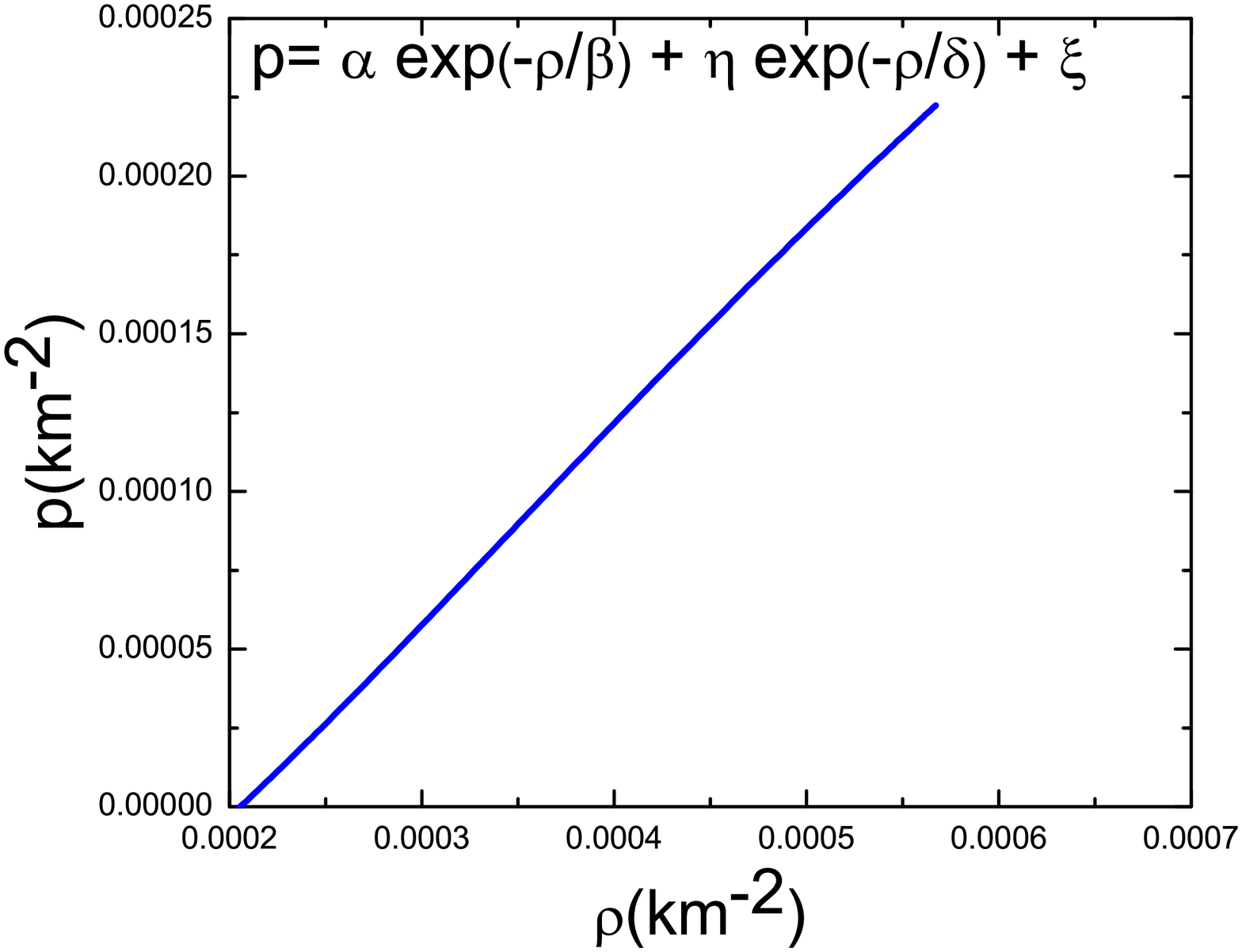}
        \caption{Possible pressure ($p$)-density ($\rho$) relation (EOS) at the stellar interior taking a=0.0016 $km^{-2}$, C=1.133,
        where $\alpha, \beta, \eta, \delta, \xi$ are constants and all are in units of $km^{-2}$.}
    \label{fig:11}
\end{figure}

It is to be noted here that the model described by Heint IIa~(\cite{Heintzmann1969}) is useful to study both neutron and strange stars depending upon the choice of the metric parameter $a$, $C$ (Kalam et al.~\cite{Kalam2016}; Kalam et al.~\cite{Kalam2017}). In this article, we have investigated that whether the same Heint IIa metric is capable to explain the compact stars within low-mass X-ray binaries or not. For which, another demonstrated the physical behavior of the six compact stars within the low-mass X-ray binary (LMXB) namely Cyg X-2, V395 Carinae/2S 0921-630, XTE J2123-058, X1822-371 (V691 CrA), 4U 1820-30 and GR Mus (XB 1254-690) by considering isotropic pressure in nature. Here we have also merged the previously cosmological constant $\Lambda$ in the Einstein's field equation in favour of study the stellar construction. Effectively, we obtained an analytical solution fot the fluid sphere another really interesting attach to diverse physical property, which are as follows:
\begin{enumerate}
\item[(i)] In our model at the interior of the compact stars density and pressure well funtion (positive definite at the centre) (Fig.~1). It is to be mentioned here that pressure and density are both maximum at the origin and interestingly pressure fall to zero (monotonically decreasing) towards the boundary while density does not.
Therefore it is justified to designated these compact stars as strange stars therein the surface density does not vanishes in place of the neutron stars dissimilar the surface density vanishes at the boundary.
Here, we assume the values of constants ($a$, $C$) in the metric and $\Lambda$ in suchlike that pressure must dissolve at the boundary. By assuming of the constant's values $a$, $C$ and $\Lambda$, we calculate the central density, $\rho_{0}$ as $567 \times 10^{-6} km^{-2} (7.651 \times 10^{14} gm/cm^3)$ and central pressure, $p_{0}$ as $2224.51 \times 10^{-7} km^{-2} (5.557 \times 10^{35} dyne/cm^2)$ (Table~1).
It satisfies energy conditions, TOV equation and Herrera's stability condition. It is also stable regard to infinitesimal radial perturbations. From mass function (equation~8), all desired inside properties of a compact star be possile to evaluated which satisfies Buchdahl mass-radius relation ($\frac{ 2M}{R} < \frac{8}{9}$) (Figs.~4, 5(left)). The surface redshift in respect of compact stars are found under the standard measure ($Z_{s}\leq 0.85$) that is favourable (Fig.~5(right)) (Haensel et al.~\cite{Haensel2000}). We estimated the EOS and that would be like $ p = \alpha e^{(-\rho/\beta )} + \eta e^{(-\rho/\delta )} + \xi $ whereinto $\alpha, \beta,~ \eta,~ \delta, ~\xi$ are constants and theirs unit of $km^{-2}$. Fig.~7 indicates that a stiff equation of state ($\ddot{O}$zel~(\cite{Ozel2006}); Lai \& Xu~(\cite{Lai2009}) and Guo et al.~(\cite{GUO2014})) rather be a soft equation of state.

\item[(ii)] From our mass function graph Fig.~6, equation (9) and equation (10), we obtain the radii, compactness and surface red-shift of six compact stars within the low-mass X-ray binary (LMXB) as like Cyg X-2, V395 Carinae/2S 0921-630, XTE J2123-058, X1822-371 (V691 CrA), 4U 1820-30 and GR Mus (XB 1254-690). The detail comparison chart are shown in Table~2.
\end{enumerate}

It is to be mentioned here that we actually considering  Heint IIa metric with de-Sitter spacetime to describe the compact stars within low-mass X-ray binaries where inlaid metric parameters $a$, $C$  are assess by computing all modes of necessary situations. When metric parameters values are known, the EOS additionally the central density are settled. In general, the mass-radius curve are considered under a conferred equation of state for different values of central density; with a definite value of the central density, the mass and radius of a compact star are settled. In spite of our model is diverse and theoretically attractive. According to our model, six compact stars within the low-mass X-ray binary (LMXB) namely Cyg X-2, V395 Carinae/2S 0921-630, XTE J2123-058, X1822-371 (V691 CrA), 4U 1820-30 and GR Mus (XB 1254-690) derive the identical values of $a$, $C$ and therefore the identical central density and the identical equation of state. Further interestingly in our stellar model, if we begin out of the center by a particular central density, the construction of a compact star be possible determined with preventing on any radius whereinto pressure arrive to zero.

Therefore, our conclusion is that we may find useful relativistic model in the sake of compact stars within low-mass X-ray binaries by suitable choice of the values of the metric parameters $a$, $C$ in the metric given by Heint IIa~(\cite{Heintzmann1969}).

\section*{Acknowledgments}
MK thankful acknowledge the help from IUCAA, Pune, India for giving research advantage and warm hospitality under Visiting Associateship programme. MK also would like to thank TWAS-UNESCO lordship and IASBS, Iran because of giving research advantage and liven hospitality under Associateship programme, whereinto a section of this work was conducted. Rabiul Islam would like to thank UGC for financial support under MANF scheme.

\label{lastpage}


\begin{thebibliography}{99}
\bibitem[2007]{Abreu2007} Abreu H., Hernandez H., Nunez L. A., 2007, Class. Quantum. Grav., 24, 4631
\bibitem[1986]{Alcock1986} Alcock C., Farhi E., Olinto A., 1986, Astrophys. J., 310, 261
\bibitem[1966]{Bardeen1966} Bardeen J. M., Thorne K. S., Meltzer D. W., 1966, Astrophys. J., 145, 505
\bibitem[2007]{Barnes2007} Barnes A. D., {\em et al.} 2007, MNRAS, 380, 1182
\bibitem[2017]{Bhar2017} Bhar P., Singh K. N., Sarkar N., Rahaman F., 2017, Eur. Phys. J. C, 77, 9
\bibitem[2005]{Bohmer2005} B$\ddot{O}$hmer C. G., Harko T., 2005, Phys. Lett. B, 630, 73
\bibitem[2005a]{Bohmer2005a} B$\ddot{O}$hmer C. G., Harko T., 2005a, Phys. Rev. D, 71, 084026
\bibitem[2016]{Bordbar2016} Bordbar G. H., Hendi S. H., Eslam Panah B., 2016, Eur. Phys. J. Plus, 131, 315
\bibitem[2006]{Bronnikov2006} Bronnikov K., Fabris J. C., 2006, Phys. Rev. Lett., 96, 251101
\bibitem[1959]{Buchdahl1959} Buchdahl, H. A., 1959, Phys. Rev., 116, 1027
\bibitem[2010]{Casares2010} Casares J., Hern´andez J. I. G., Israelian G., Rebolo R., 2010, MNRAS, 401, 2517
\bibitem[1964]{Chandrasekhar1964} Chandrasekhar S., 1964, Astrophys. J., 140, 417
\bibitem[1990]{Chen1990} Chen W., Wu Y. S., 1990, Phys. Rev. D, 41, 695
\bibitem[2016]{Dayanandan2016} Dayanandan B., Maurya S. K., Gupta Y. K., Smitha T. T., 2016, Astrophys Space Sci., 361, 160
\bibitem[1998]{Dey1998} Dey M., Bombaci I., Dey J., Ray S., Samanta B. C., 1998, Phys. Lett. B, 438, 123
\bibitem[2014]{Drago2014} Drago A., Lavagno A., Pagliara G., 2014, Phys. Rev. D, 89, 043014
\bibitem[2002]{Dymnikova2002} Dymnikova I., 2002, Class. Quantum. Gravit., 19, 725
\bibitem[2003]{Dymnikova2003} Dymnikova I., 2003, Int. J. Mod. Phys. D, 12, 1015
\bibitem[2007]{Egeland2007} Egeland E., 2007, {\it Compact Star}, Trondheim, Norway
\bibitem[1984]{Farhi1984} Farhi E., Jaffe R. L., 1984, Phys. Rev. D, 30, 2379
\bibitem[2003]{Gelino2003} Gelino D. M., Tomsick J. A., Heindl W. A., 2003, Bull. Am. Astron. Soc., 34, 1199
\bibitem[2014]{GUO2014} GUO Yan-Jun, LAI Xiao-Yu, XU Ren-Xin, 2014, Chinese Physics C, 38, 5
\bibitem[2010a]{Guver2010a} G$\ddot{u}$ver T., $\ddot{O}$zel F., Cabrera-Lavers A., 2010a, Astrophys. J., 712, 964
\bibitem[2010b]{Guver2010b} G$\ddot{u}$ver T., Wroblewski P., Camarota L., $\ddot{O}$zel F., 2010b, Astrophys. J., 719, 1807
\bibitem[1986]{Haensel1986} Haensel P., Zdunik Z. L., Schaeffer R., 1986, Astron. Astrophys., 160, 121
\bibitem[2000]{Haensel2000} Haensel P., Lasopa J. P., Zdunik J. L., 2000, Nucl. Phys. Proc. Suppl., 80, 1110
\bibitem[1992]{Heap1992} Heap S. R., Corcoran M. F., 1992, APJ, 387, 340
\bibitem[1969]{Heintzmann1969} Heintzmann H., 1969, Z. Phys., 228, 489
\bibitem[1992]{Herrera1992} Herrera L., 1992, Phys. Lett. A, 165, 206
\bibitem[2012]{Hossein2012} Hossein Sk. M., Rahaman F., Naskar J., Kalam M., Ray S., 2012, Int. J. Mod. Phys. D, 21, 1250088
\bibitem[2017]{Jafry2017} Jafry M. A. K., {\em et al.} 2017, Astrophys. Space Sci., 362, no.10, 188
\bibitem[2012]{Kalam2012} Kalam M., Rahaman F., Ray S., Hossein S. M., Karar I., Naskar J., 2012, Eur. Phys. J. C, 72, 2248
\bibitem[2013a]{Kalam2013a} Kalam M., Usmani A. A., Rahaman F., Hossein S. M., Karar I., Sharma R., 2013a, Int. J. Theor. Phys., 52, 3319
\bibitem[2013b]{Kalam2013b} Kalam M., Rahaman F., Hossein S. M., Ray S., 2013b, Eur. Phys. J. C, 73, 2409
\bibitem[2014a]{Kalam2014a} Kalam M., Rahaman F., Molla S., Jafry M. A. K., Hossein S. M., 2014a, Eur. Phys. J. C, 74, 2971
\bibitem[2014b]{Kalam2014b} Kalam M., Rahaman F., Molla S., Hossein S. M., 2014b, Astrophys. Space Sci., 349, 865
\bibitem[2016]{Kalam2016} Kalam M., {\em et al.} 2016, Mod. Phys. Lett. A, 31, No. 40, 1650219
\bibitem[2017]{Kalam2017} Kalam  M., {\em et al.} 2017, Mod. Phys. Lett. A, 32, No. 4, 1750012
\bibitem[1988]{Knutsen1988} Knutsen H., 1988, MNRAS, 232, 163
\bibitem[2009]{Lai2009} Lai X. Y., Xu R. X., 2009, MNRAS, 398, L31-L35
\bibitem[2005]{lattimer2005} Lattimer J. M., Prakash M., 2005, Phys. Rev. Lett., 94, 111101
\bibitem[2007]{lattimer2007} Lattimer J. M., Prakash M., 2007, Phys. Rep., 442, 109
\bibitem[2006]{Lobo2006} Lobo F., 2006, Class. Quantum. Grav., 23, 1525
\bibitem[2014]{Maharaj2014} Maharaj S. D., Sunzub J. M., Ray S., 2014, Eur. Phys. J. Plus, 129, 3
\bibitem[2000]{MaK2000} MaK M. K., {\em et al.} 2000, Mod. Phys. Lett. A, 15, 2153
\bibitem[2013]{Mak2013} Mak M. K., Harko T., 2013, Eur. Phys. J. C, 73, 2585
\bibitem[2016]{Maurya2016} Maurya S. K., Gupta Y. K., Dayanandan B., Ray S., 2016, Eur. Phys. J. C, 76, 266
\bibitem[2005]{Munoz2005} Mu\~{n}oz-Darius T., {\em et al.} 2005, Astrophys. J., 635, 502
\bibitem[1991]{Narlikar1991} Narlikar J. V., Pecker J. C., Vigier J. P., 1991, J. Astrophys. Astr., 12, 7
\bibitem[2015]{Ngubelanga2015} Ngubelanga S., Maharaj S. D., Ray S., 2015, Astrophys Space Sci., 357, 74
\bibitem[1999]{Orosz1999} Orosz J. A., Kuulkers E., 1999, MNRAS, 305, 132
\bibitem[2006]{Ozel2006} $\ddot{O}$zel F., 2006, Nature, 441, 1115
\bibitem[2009a]{Ozel2009a} $\ddot{O}$zel F., G$\ddot{u}$ver T., Psaitis D., 2009a, Astrophys. J., 693, 1775
\bibitem[2009b]{Ozel2009b} $\ddot{O}$zel F., Psaitis D., 2009b, Phys. Rev. D, 80, 103003
\bibitem[2010]{Ozel2010} $\ddot{O}$zel F., Baym G., G$\ddot{u}$ver T., 2010, Phys. Rev. D, 82, 101301
\bibitem[2014]{Pant2014} Pant N., Pradhan N., Murad M. H., 2014, Int. J. Theor. Phys., 53, 11
\bibitem[2015]{Paul2015} Paul B. C., Chattopadhyay P. K., Karmakar S., 2015, Astrophys. Space Sci., 356, 327
\bibitem[1998]{Perlmutter1998} Perlmutter S., {\em et al.} 1998, Nature, 391, 51
\bibitem[2012a]{Rahaman2012a} Rahaman F., Maulick R., Yadav A. K., Ray S., Sharma R., 2012a, Gen. Relativ. Gravit., 44, 107
\bibitem[2012b]{Rahaman2012b} Rahaman F., Sharma R., Ray S., Maulick R., Karar I., 2012b, Eur. Phys. J. C, 72, 2071
\bibitem[1993]{Ray1993} Ray S., Ray D., 1993, Astrophys. Space Sci., 203, 211
\bibitem[2004]{Riess2004} Riess A. G., {\em et al.} 2004, Astrophys. J., 607, 665
\bibitem[2007]{Steeghs2007} Steeghs D., Jonker P. G., 2007, Astrophys. J., 669, L85
\bibitem[1997]{Stickland1997} Stickland D., Lloyd C., Radzuin-Woodham A., 1997, MNRAS, 286, L21
\bibitem[1995]{Van1995} Van Kerkwijk J. H., van Paradijis J., Zuiderwijk E. J., 1995, A$\&$A, 303, 497
\bibitem[1967]{Zel'dovich1967} Zel'dovich Y. B., 1967, JETP Lett., 6, 316
\bibitem[1968]{Zel'dovich1968} Zel'dovich Y. B., 1968, Sov. Phys. Usp., 11, 381


\end{thebibliography}
\end{document}